\documentclass[referee,traditabstract]{aa}
\usepackage{graphics}
\usepackage{dcolumn}
\usepackage[reqno]{amsmath}
\usepackage{amsfonts}
\usepackage{amssymb}
\usepackage{bm}
\usepackage{hyperref}
\usepackage{url}
\usepackage{subfigure}
\usepackage{natbib}
\usepackage{txfonts}
\usepackage{color}
\begin{document}

\newcommand{\changeref}[1]{\textbf{#1}}
\newcommand{\changeJ}[1]{\textcolor{blue}{#1}}
\newcommand{\pot}[2]{{#1\times 10^{#2}}}

\newcommand{\changeR}[1]{\textcolor{red}{#1}}
\newcommand{\changeJII}[1]{\textcolor{red}{#1}}
\newcommand{\Tnew}{{{T_{\rm new}}}}
\newcommand{\TCMB}{{{T_{\rm CMB}}}}
\newcommand{\ECMB}{{{E_{\rm CMB}}}}
\newcommand{\Eg}{{{E_{\gamma}}}}
\newcommand{\NCMB}{{{N_{\rm CMB}}}}
\newcommand{\SCMB}{{{S_{\rm CMB}}}}
\newcommand{\Spl}{{{S_{\rm pl}}}}
\newcommand{\Te}{{{T_{\rm e}}}}
\newcommand{\Teq}{{{T^{\rm eq}_{\rm e}}}}
\newcommand{\Ti}{{{T_{\rm i}}}}
\newcommand{\nB}{{{n_{\rm B}}}}
\newcommand{\nS}{{{n_{\rm s}}}}
\newcommand{\ve}{{{\rm v}}}
\newcommand{\Teff}{{{T_{\rm eff}}}}

\newcommand{\id}{{{\rm d}}}
\newcommand{\aR}{{{a_{\rm R}}}}
\newcommand{\bR}{{{b_{\rm R}}}}
\newcommand{\neb}{{{n_{\rm eb}}}}
\newcommand{\kB}{{{k_{\rm B}}}}
\newcommand{\EB}{{{E_{\rm B}}}}
\newcommand{\zmin}{{{z_{\rm min}}}}
\newcommand{\zmax}{{{z_{\rm max}}}}
\newcommand{\YBEC}{{{Y_{\rm BEC}}}}
\newcommand{\YSZ}{{{Y_{\rm SZ}}}}
\newcommand{\rhob}{{{\rho_{\rm b}}}}
\newcommand{\Ne}{{{n_{\rm e}}}}
\newcommand{\sigT}{{{\sigma_{\rm T}}}}
\newcommand{\me}{{{m_{\rm e}}}}
\newcommand{\npl}{{{n_{\rm Pl}}}}

\newcommand{\kD}{{{{k_{\rm D}}}}}
\newcommand{\KC}{{{{K_{\rm C}}}}}
\newcommand{\KdC}{{{{K_{\rm dC}}}}}
\newcommand{\Kbr}{{{{K_{\rm br}}}}}
\newcommand{\zdC}{{{{z_{\rm dC}}}}}
\newcommand{\zbr}{{{{z_{\rm br}}}}}
\newcommand{\aC}{{{{a_{\rm C}}}}}
\newcommand{\adC}{{{{a_{\rm dC}}}}}
\newcommand{\abr}{{{{a_{\rm br}}}}}
\newcommand{\gdC}{{{{g_{\rm dC}}}}}
\newcommand{\gbr}{{{{g_{\rm br}}}}}
\newcommand{\gff}{{{{g_{\rm ff}}}}}
\newcommand{\xe}{{{{x_{\rm e}}}}}
\newcommand{\alphafs}{{{{\alpha_{\rm fs}}}}}
\newcommand{\YHe}{{{{Y_{\rm He}}}}}
\newcommand{\SE}{{{{S_{\rm E}}}}}
\newcommand{\SN}{{{{S_{\rm N}}}}}
\newcommand{\muc}{{{{\mu_{\rm c}}}}}
\newcommand{\xc}{{{{x_{\rm c}}}}}
\newcommand{\xH}{{{{x_{\rm H}}}}}
\newcommand{\mT}{{{{\mathcal{T}}}}}
\newcommand{\Ob}{{{{\Omega_{\rm b}}}}}
\newcommand{\Or}{{{{\Omega_{\rm r}}}}}

\title{Mixing of blackbodies: entropy production and dissipation of sound
  waves in the early Universe}

\author{Rishi Khatri\inst{\ref{inst1}}
\and
Rashid A. Sunyaev\inst{\ref{inst1}{,}\ref{inst2}{,}\ref{inst3}}
\and
Jens Chluba\inst{\ref{inst4}}
}

\institute{Max Planck Institut f\"{u}r Astrophysik,
  Karl-Schwarzschild-Str. 1, 85741 Garching, Germany\\
\email{khatri@mpa-garching.mpg.de}\label{inst1}
\and
 Space Research Institute, Russian Academy of Sciences, Profsoyuznaya
 84/32, 117997 Moscow, Russia \label{inst2}
\and
Institute for Advanced Study, Einstein Drive, Princeton, New Jersey 08540, USA\label{inst3}
\and
Canadian Institute for Theoretical Astrophysics, 60 St George Street, Toronto, ON M5S 3H8, Canada\label{inst4}
}
\date{\today}

\abstract
{
Mixing of  blackbodies with different temperatures creates a spectral distortion
which,  at lowest order, is a
$y$-type distortion, indistinguishable from the thermal $y$-type distortion
produced by the scattering of CMB photons by hot electrons residing in
clusters of galaxies. This process occurs in the
radiation-pressure dominated early Universe, when the primordial
  perturbations excite standing sound waves on entering the sound horizon.  Photons from  different
phases of the sound waves, having different temperatures, diffuse through the electron-baryon plasma and mix
together. This diffusion, with the length defined
by Thomson scattering, dissipates sound waves and creates
spectral distortions in the CMB. Of the total dissipated energy, $2/3$ raises the average temperature of the blackbody part
of spectrum, while $1/3$ creates a distortion of $y$-type. It is well known
that at redshifts $10^5\lesssim z\lesssim 2\times 10^6$, comptonization
rapidly transforms $y$-distortions into a Bose-Einstein spectrum. The
chemical potential of the Bose-Einstein spectrum is again $1/3$ the value
we would get if all the dissipated energy was injected into a blackbody
spectrum but no extra
photons were added. We study the mixing of blackbody spectra, emphasizing the thermodynamic point of
view, and identifying spectral distortions with entropy creation. This allows us to obtain the main results connected with the dissipation
of sound waves in the early Universe in a very simple way. We also show
that mixing of blackbodies in general, and dissipation of sound waves in
particular, leads to creation of entropy.
}
\titlerunning{Mixing of blackbodies}

\maketitle
\section{Introduction}
The Cosmic microwave background (CMB) has a spectrum which is close to a blackbody 
  with very high
 precision. COBE/FIRAS \citep{cobe} constrained possible departures from a perfect
blackbody to be $\lesssim 10^{-5}$ for $y$-type distortions, and
the   chemical
potential was limited to $|\mu|\lesssim 9\times 10^{-5}$. This observation immediately has an important
consequence: there was a time in the history of the Universe when matter
and radiation were in 
complete thermodynamic equilibrium with each other and any energy release
at $z\lesssim 2\times 10^6$ was small.

In addition to being an almost perfect blackbody, the CMB is also
isotropic, with anisotropies of less than $10^{-3}$. The dominant component of
the anisotropy is  the dipole  caused by our peculiar motion. The
anisotropies other than the dipole are at the level of $10^{-5}$ and
originate from  the
primordial fluctuations imprinted in the initial conditions of the present expanding Universe. The radiation field seen
by any observer (or electrons in the hot gas in a  cluster of galaxies/early Universe) will
thus consist of blackbodies  having  temperature that differs as a
function of observation direction. Scattering by electrons,
averaging of the smaller scales within the beam of
the telescope  or explicit
averaging by a cosmologist, will thus inevitably mix these blackbodies
together. Mixing of blackbodies was first
 studied in  detail by \citep{Zeldovich1972} who showed that
it creates a $y$-type distortion,  
  indistinguishable from the $y$-type distortion created by interaction of the
 blackbody photons with hotter electrons \citep{zs1969}. \citet{cs2004}
 proved that at lowest order, this result is valid for arbitrary 
  temperature distributions and not just the Gaussian ones. Comptonization of
 this initial $y$-type distortion 
 converts it into a Bose-Einstein spectrum or a $\mu$-type
distortion \citep{zs1969,cks2012} for $y\gtrsim 1$, where the Compton $y$-parameter is defined as
the following integral over time $t$: 
$y=\int  \Ne \sigT c \frac{\kB\Te}{\me c^2}\id t$, $\Ne$ is the
electron number density, $\sigT$ is the Thomson cross section, $c$ is the speed
of light, $\kB$ is the Boltzmann's constant, $\Te$ is the electron temperature,
and $\me$ is the mass of electron \citep{k1956,zs1969}. The purpose of this
paper is to study the  mixing of blackbodies from a statistical
physics point
of view. Amazingly, using basic thermodynamic relations, we arrive at the main results
connected with the dissipation of sound waves in the early Universe due to
shear viscosity and thermal conduction in a very simple manner. Some of these
aspects are also discussed in \citet{cks2012}.

\section{Entropy of a Bose gas}
The entropy density of a general distribution of a Bose gas (photons),
which may or may not be in equilibrium, is given by \citep{llstats}
\begin{align}
S&=\int \id\nu \frac{8\pi
  \nu^2}{c^3}\left[(1+n)\ln(1+n)-n\ln(n)\right]\nonumber\\
&=8\pi\left(\frac{\kB T
  }{h c}\right)^3\int  x^2\left[(1+n)\ln(1+n)-n\ln(n)\right]~\id x
\end{align}
where $\frac{g d^3p}{h^3}=\frac{8\pi \nu^2\id \nu}{c^3}$ is the  density of available
states in the frequency interval $\id\nu$, and for photons the degeneracy
$g=2$,
$h$ is Planck's
constant, $n=\frac{c^3 E_{\nu}}{8\pi h \nu^3}$ is the photon occupation
number and $E_{\nu}$ is the energy density of photons per unit frequency. We have changed variables to dimensionless frequency
$x=h\nu/(\kB T)$ defined with respect to a reference temperature $T$ in the
second line.
For a blackbody spectrum at temperature $T$ with $n=\npl(x)\equiv 1/(e^x-1)$,
we obtain $\Spl=\frac{32 \pi^5\kB^3 T^3}{45 c^3 h^3}\equiv \frac{4}{3}\aR
T^3/\kB$, where the last line defines the radiation constant $\aR$.

Now let us add a small spectral distortion $\delta n$ to the Planck
spectrum so that the total occupation number is $n=\npl(1+\delta n/\npl)$,
where $\delta n/\npl<<1$. Expanding the expression for entropy up to first
order in $\delta n/\npl$ and ignoring the higher order terms, we get
$S=\Spl+\delta S$, where $\delta S$ is the entropy density
added/subtracted due to
the spectral distortion and is given by,
\begin{align}
\delta S&\approx 8\pi\left(\frac{\kB T
  }{h c}\right)^3\int  x^2\left[\ln(1+\npl)-\ln(\npl)\right]\delta
n~~\id x\nonumber\\
&=8\pi\left(\frac{\kB T
  }{h c}\right)^3\int  x^3\delta
n~~\id x\equiv\frac{\delta Q}{\kB T},\label{noneqent}
\end{align}
where $\delta Q$ is the energy density in the distortion. Thus for small distortions,
the classic thermodynamical formula for the change in the equilibrium entropy due to addition/subtraction
of energy remains valid even in non-equilibrium. We note in particular that, at first order in
distortions, the entropy does not
depend on the shape of the distortion but only on the total energy density in the
distortion. This means  that the process of comptonization, 
which converts an initial  $y$-type distortion to a $\mu-$type distortion, in the
absence of any additional heating/cooling, does not
change the entropy of the photon gas at first order.
 The factor of $\kB$ in Eq. \eqref{noneqent} just converts
temperature from Kelvin to energy units to make entropy dimensionless. 
As an example, we can calculate the entropy produced during cosmological
recombination,  when $\sim 5$ energetic recombination photons are produced per
hydrogen atom \citep{cs2006b}. Adding up the energy in the recombination
spectrum \citep{rcs2008}, we get $\delta S_{\rm recombination}\sim \delta
Q_{\rm recombination}/(\kB \TCMB)\sim 9\times 10^{-6}~/ {\rm
  cm^3}$ today, compared with entropy density of $1478~/ {\rm
  cm^3}$ in the blackbody part of CMB with temperature $\TCMB=2.725~{\rm K}$.

In
the rest of the paper we will measure frequency and temperature in energy
units and set $c=h=\kB=1$ unless explicitly specified in definitions of
other constants.
 
\section{Mixing of blackbodies}
Since the blackbody radiation is an equilibrium distribution for photons it is
described by a single parameter, the temperature $T$. In addition, if we
specify a second thermodynamic  quantity such as volume $V$, we can
calculate any other thermodynamic quantity such as entropy or internal
energy. The spectrum is described by the well known Planck function, and
the energy density per unit frequency and total energy density integrated
over frequency can be written as,
\begin{align}
E_{\nu}&=8\pi \nu^3\frac{1}{e^{\frac{\nu}{ T}}-1}\nonumber\\
E&=a_RT^4.
\end{align}
 The number density of photons is given by $N=b_RT^3$ where 
$\bR=\frac{16 \pi\kB^3\zeta(3)}{ c^3 h^3}$, $\zeta$ is the Riemann zeta
function with $\zeta(3)\approx 1.20206$. We can also calculate the entropy
density and it is given by $S=4E/(3 T)=(4/3) a_RT^3$.
For the CMB with $\TCMB=2.725{\rm ~K}\pm 1{\rm ~mK}$ \citep{fm2002} we have energy density 
$\ECMB=0.26{\rm  ~eV/cm^3}$, number density $\NCMB=411/{\rm cm^3}$ and entropy density
$\SCMB=1478/{\rm cm^3}$. For simplicity, we will consider mixing of two
blackbodies below, but the derivation and the results are trivially generalized to an ensemble
of arbitrary number of blackbodies by just replacing the average of
quantities over two blackbodies with the appropriate average over the whole
ensemble.

\subsection{$y$-type distortion}
\begin{figure}
\resizebox{\hsize}{!}{\includegraphics{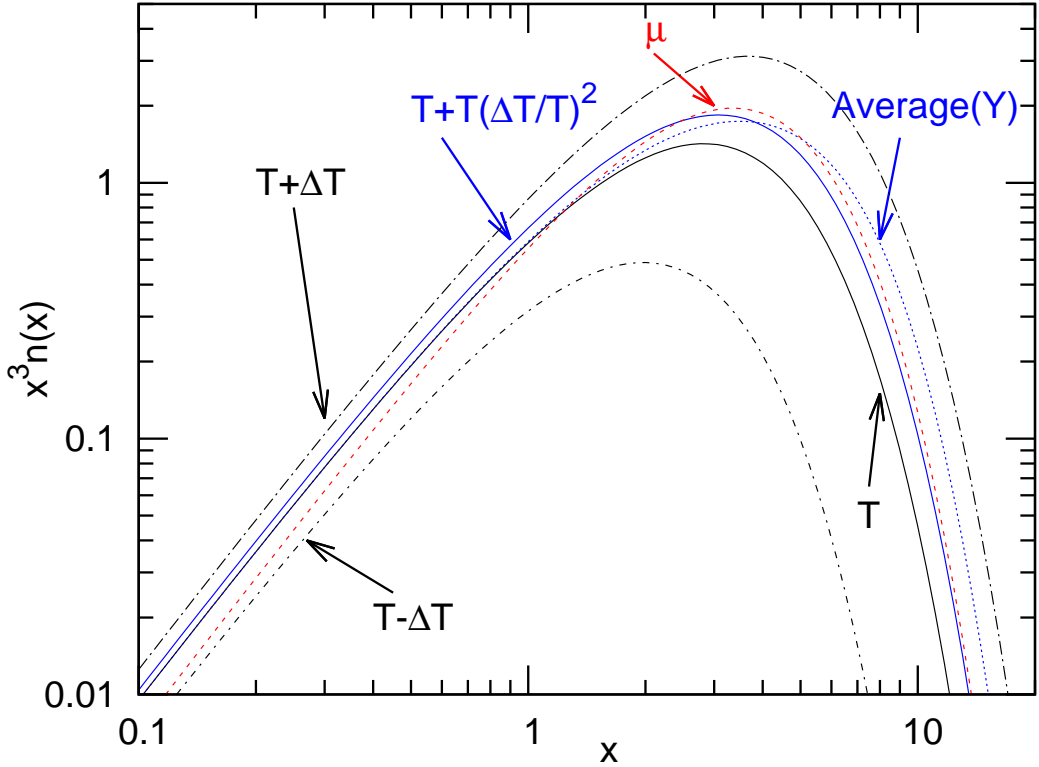}}\caption{\label{yfig} Average
  of blackbodies of temperature $T+\Delta T$ and $T-\Delta T$ creates a new
  spectrum marked 'Average(Y)' which is different from blackbody spectrum
  marked with temperature $T$. The average spectrum is just the usual $y$-
  type distortion with respect to a new blackbody at temperature
  $T[1+(\Delta T/T)^2]$, with the two crossing at $x=3.83$. At redshifts
  $z\gtrsim 10^5$ the average spectrum will comptonize to Bose-Einstein
  spectrum marked $\mu$ above. All
  three spectra, $T[1+(\Delta T/T)^2]$, Average(Y), and $\mu$, have the same number density of photons. Average/$y$-type spectrum and
  Bose-Einstein/$\mu$-type spectrum also have the same energy density which is greater
than the energy density in the blackbody spectrum  with temperature $T[1+(\Delta T/T)^2]$ by
$1/3$ of the  initial energy density excess over that of blackbody with
temperature $T$. We have used linear order formulae to calculate the $y$
and $\mu$ distortions in the figure but used a large value of $\Delta T/T$ to make
differences between different curves visible.
}
\end{figure}

\begin{figure}
\resizebox{\hsize}{!}{\includegraphics{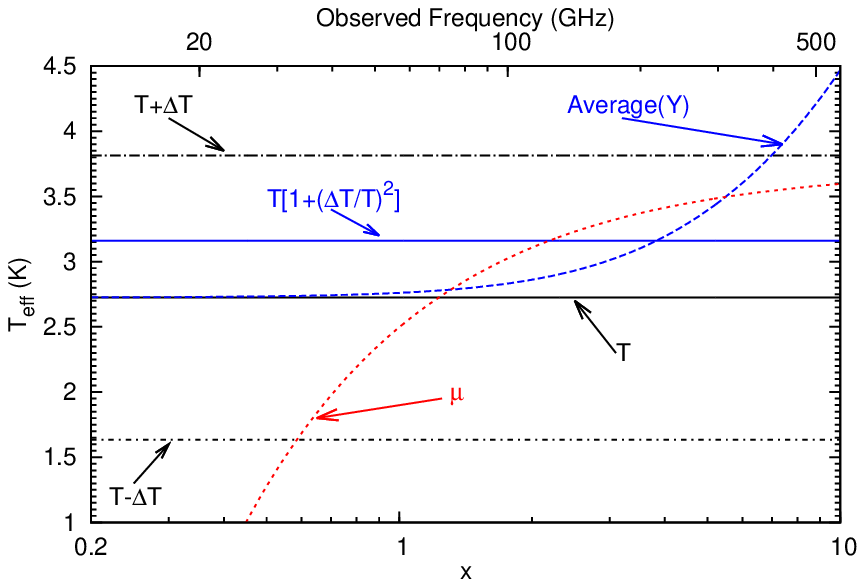}}\caption{\label{tefffig}
  Same as Fig. \ref{yfig} but effective temperature $T_{\rm eff}(\nu)$ defined
  by $n(\nu)=1/({e^{\nu/T_{\rm eff}(\nu)}-1})$ is plotted. We have used linear order formulae to calculate the $y$
and $\mu$ distortions in the figure but used a large value of $\Delta T/T=0.4$ to make
differences between different curves visible. Average temperature
$T=2.725~{\rm K}$. For Bose-Einstein spectrum we thus have, at linear order,
$T_{\rm eff}(x)=T_{\rm
BE}-T\mu/x$. For a general distortion $\delta n/n$ (including $y$-type ) we have
$T_{\rm eff}(x)=T+T\left[(1-e^{-x})/x\right]\delta n/n$. }
\end{figure}

If we mix blackbody spectra with different temperature, the resultant
spectrum is not blackbody and at lowest order the distortion is given by a
$y$-type spectrum \citep{Zeldovich1972}.\footnote{We discuss different
    processes which can lead to the mixing of blackbodies in
    the following sections.} This can be seen immediately by
Taylor expanding the photon intensity or equivalently the occupation number $n$
of a blackbody  at a temperature $T+\Delta T$ about the average
temperature $T$ and take the average (ensemble or spatial) keeping terms
up to second order in $\Delta T/T$,
\begin{align}
&\left<\npl(T+\Delta T)\right>\equiv \left<\frac{1}{e^{\frac{\nu}{(T+\Delta T)}}-1}\right>\nonumber\\
\approx &\left<\npl(T)+\ln\left[1+\frac{\Delta T}{T}\right]\frac{\partial \npl}{\partial
  \ln[T]}+\frac{1}{2}\left(\ln\left[1+\frac{\Delta T}{T}\right]\right)^2\frac{\partial^2 \npl}{\partial
  (\ln[T])^2}\right>\nonumber\\
=&\npl(T)+\left(\left<\frac{\Delta T}{T}\right>+\left<\left(\frac{\Delta
        T}{T}\right)^2\right>\right)T\frac{\partial \npl}{\partial
  T}+\frac{1}{2}\left<\left(\frac{\Delta
      T}{T}\right)^2\right>T^4\frac{\partial }{\partial
  T}\frac{1}{T^2}\frac{\partial \npl}{\partial T}\nonumber\\
=&\npl\left[T\left(1+\left<\left(\frac{\Delta
        T}{T}\right)^2\right>\right)\right]+\frac{1}{2}Y(x)\left<\left(\frac{\Delta
        T}{T}\right)^2\right>,\label{taylor}
\end{align}
where we have used $\left<\frac{\Delta T}{T}\right>\equiv 0$. The first
term in the last line is simply a blackbody with temperature $\Tnew=T[1+(\Delta T/T)^2] > T$, and 
\begin{align}
Y(x)=\frac{xe^x}{(e^x-1)^2}\left(x\frac{e^x+1}{e^x-1}-4\right)
\end{align}
is
the $y$-type spectrum with $x=\nu/T$ \citep{zs1969} with the magnitude of
distortion given by $y_Y=\frac{1}{2}\left<\left(\frac{\Delta
        T}{T}\right)^2\right>$. The average  spectrum with
  $y$-type distortion  for
  average of two blackbodies with temperature $T\pm \Delta T$ is shown
in Fig. \ref{yfig}. Figure \ref{tefffig} shows the same spectra but in
terms of the effective temperature
defined by $n(\nu)=1/({e^{\nu/T_{\rm eff}(\nu)}-1})$, which for small distortions
can be written, at linear order, as
$T_{\rm eff}(x)=T+T\left[(1-e^{-x})/x\right]\delta n/n$. We note that in the
Rayleigh-Jeans limit, intensity is proportional to temperature, and thus the
average of blackbody spectra just gives the spectrum at average
temperature $T$, and $T_{\rm eff}\rightarrow T$. An important property of $y$-type distortion is that it
represents pure redistribution of photons in the spectrum due to addition
of energy but conserves photon number. Thus $\int  x^2 Y(x)~\id x=0$ and the
above mentioned decomposition of the spectrum into a blackbody part and
$y$-distortion part is unique and independent of gauge/reference frame, at
order $\delta n/n$, in
the following sense: The constraint $\int  x^2 \delta n(x)~\id x=0$ on the
spectral distortion $\delta n$ fixes the reference temperature of the
blackbody part of the spectrum used to define the variable $x$ and $\delta
n$ as a function of the variable $x$ is gauge
independent.\footnote{In fact $\delta n(x)$ is also invariant to
  change in the reference temperature happening in the mixing of
  blackbodies, $\delta n(x)=\delta n(x_{\rm {new}})+\mathcal{O}([\delta
  n/n]^2)$, where $x_{\rm new}=\nu/\Tnew$.}

Without loss of generality, let us consider the superposition of two blackbody spectra
with temperatures $T_1=T+\Delta T$ and $T_2=T-\Delta T$, with average
temperature $T$.
The average initial energy density, number density and entropy density  of the two blackbodies is 
\begin{align}
E_{\rm initial}&=\frac{1}{2}\aR(T_1^4+T_2^4)\approx \aR
T^4\left[1+6\left(\frac{\Delta T}{T}\right)^2\right]>\aR T^4\nonumber\\
N_{\rm initial}&=\frac{1}{2}\bR(T_1^3+T_2^3)\approx \bR
T^3\left[1+3\left(\frac{\Delta T}{T}\right)^2\right]>\bR T^3\nonumber\\
S_{\rm initial}&=\frac{1}{2}\frac{4\aR}{3}(T_1^3+T_2^3)\approx \frac{4\aR}{3}
T^3\left[1+3\left(\frac{\Delta T}{T}\right)^2\right]>\frac{4\aR}{3} T^3 \label{initialeq}
\end{align}
We can calculate the final temperature of a blackbody having the same
number density of photons as the initial average $N_{\rm initial}$.
\begin{align}
T_{\rm final}=\left(\frac{N_{\rm initial}}{\bR}\right)^{1/3}\approx T\left[1+\left(\frac{\Delta T}{T}\right)^2\right].
\end{align}
This is exactly the temperature, $\Tnew$, of the blackbody we got by averaging the
intensity in Eq.~\eqref{taylor}. Thus \emph{all} the initial photons go into
creating a blackbody with a higher temperature. The entropy density of this
new blackbody is also identical to the initial average entropy density
$S_{\rm initial}$ because number density and entropy density have the same
$T^3$ temperature dependence. The energy density of the new blackbody is however given by,
\begin{align}
E_{\rm final}&=\aR T_{\rm final}^4 \approx \aR
T^4\left[1+4\left(\frac{\Delta T}{T}\right)^2\right]<E_{\rm initial}.
\end{align}
In fact we find
\begin{align}
E_{\rm final}-\aR T^4&=2/3\left(E_{\rm initial}-\aR
  T^4\right).
\end{align} 
This result can also be obtained directly by multiplying
Eq.~\eqref{taylor} by $\nu^3$ and integrating over frequency.
Equation~\eqref{taylor} also shows that the rest of the initial energy, equal to  
\begin{align}
1/3\left(E_{\rm initial}-\aR
  T^4\right)&=2\left(\frac{\Delta T}{T}\right)^2 \aR
  T^4 \propto\frac{1}{2}\left(\frac{\Delta T}{T}\right)^2 \!\int ~\id x \,
x^3 Y(x),
\end{align}
 goes to the
$y$-distortion with the magnitude of the distortion given by
$y_Y=\frac{1}{2}\left(\frac{\Delta T}{T}\right)^2$. These results
  were recently obtained in \citet{cks2012} using the Boltzmann equation.
It
is well known \citep{zs1969} that $y$-distortion of magnitude $y_Y$
decreases the brightness temperature of radiation in the Rayleigh-Jeans
part of the spectrum by an amount $-2y_Y$ which is equal to $\left(\frac{\Delta
    T}{T}\right)^2$ for the mixing of blackbodies considered here. But we
have also increased the magnitude of the brightness temperature by same
amount in the blackbody part of the spectrum, resulting in the brightness
temperature in the Rayleigh-Jeans part for the total spectrum which is equal to the average temperature
$T$ of the initial blackbodies, as shown in Figs. \ref{yfig} and
\ref{tefffig}.  We can calculate the additional entropy in the
  final spectrum resulting from the 
mixing of blackbodies using Eq.~\eqref{noneqent},
\begin{align}
\label{eq:DS_express}
\Delta S=\frac{\Delta E}{T}\approx 2\aR T^3 \left(\frac{\Delta T}{T}\right)^2,
\end{align}
where $\Delta E=1/3\left(E_{\rm initial}-\aR
  T^4\right)$ is the energy in spectral distortion, i.e. deviation from the
blackbody spectrum.

To summarize, mixing/averaging of blackbodies leads to a new spectrum which
has a blackbody part at a temperature which is higher than the  average
temperature of initial blackbodies by $\Delta \Tnew=T(\Delta T/T)^2$. This
new blackbody has the same entropy as
initial average entropy and the same number of photons but less energy. The
remaining  energy density appears as a $y$-type distortion, which is just a 
 redistribution of the photons of the new blackbody spectrum. The
 $y$-distortion part can be identified with the increase in the entropy of
 the system (Eq.~\eqref{eq:DS_express}), which is expected as there is an
 increase in disorder of the system.

\subsection{$\mu$-type distortion}
We saw in the previous section that it is impossible to create a blackbody
spectrum 
by mixing/averaging blackbodies of different temperatures. The reason is
that there is too much energy density for the given number density of
photons. However we can
still create a Bose-Einstein spectrum with a non zero chemical potential $\mu$, which is the spectrum we would get
if the photons could again reach equilibrium while conserving energy and
number. It is straightforward to calculate the temperature and chemical
potential of the Bose-Einstein spectrum, $n_{\rm BE}=1/(e^{\nu/T_{\rm
  BE}+\mu}-1)$,  
by equating the initial average photon number $N_{\rm initial}$
and energy density $E_{\rm initial}$
to the photon number $N_{\rm BE}$ and energy density $E_{\rm BE}$ in a Bose-Einstein spectrum, as is
done in case of heating of CMB without adding additional photons 
\citep{is1975}. In the limit of small chemical potential, $\mu\ll 1$, with
$T_{\rm BE}=\Tnew(1+t_{\rm BE})\approx
T\left(1+t_{\rm BE}+[\Delta T/T]^2\right),t_{\rm BE}<<1$
we have \citep{is1975}\footnote{Using $6\zeta(3)/I_3\approx 1.1106$ and
  $\pi^2/(3I_2)\approx 1.3684$, where $\zeta$ is the Riemann zeta function with $\zeta(3)\approx 1.20206$,
and $I_3=\int x^3\npl(x)~\id x=\pi^4/15,I_2= \int x^3\npl(x)~\id x=2\zeta(3)$.}
\begin{align}
E_{\rm BE}&\approx \aR T_{\rm BE}^4\left(1-1.1106 \mu \right)\nonumber\\
&\approx \aR
T^4\left(1+4 t_{\rm BE}+4\left(\frac{\Delta T}{T}\right)^2-1.1106 \mu
\right)\nonumber\\
=E_{\rm initial}&\approx \aR
T^4\left[1+6\left(\frac{\Delta T}{T}\right)^2\right],\nonumber\\
N_{\rm BE}&\approx  \bR T_{\rm BE}^3\left(1-1.3684 \mu \right)\nonumber\\
&\approx \bR
T^3\left(1 + 3 t_{\rm BE}+3\left(\frac{\Delta T}{T}\right)^2-1.3684 \mu
\right)\nonumber\\
=N_{\rm initial}&\approx  \bR
T^3\left[1+3\left(\frac{\Delta T}{T}\right)^2\right]
.
\end{align}
The second equation, $N_{\rm BE}=N_{\rm initial}$,  directly gives us the
relation $t_{\rm BE}=1.3684 \mu/3=\mu/2.19$. Solving the system of
equations for $t_{\rm BE}$ and $\mu$ in terms of $\Delta
T/T$, we get
\begin{align}
\mu&=1.4\left[\frac{1}{3}\left(E_{\rm initial}-\aR
  T^4\right)\right]=2.8 \left(\frac{\Delta T}{T}\right)^2\nonumber\\
t_{\rm BE}&=\frac{\mu}{2.19}=1.278\left(\frac{\Delta T}{T}\right)^2
\end{align}
This is exactly the result we will get if we add energy $\Delta E =\frac{1}{3}\left(E_{\rm initial}-\aR
  T^4\right)$ to a blackbody with temperature $\Tnew$ while
conserving the photon number. We can thus write the deviation of the
  $\mu$-type spectrum from the blackbody with temperature $\Tnew$:
\begin{align}
\frac{n_{\rm BE}-\npl(\Tnew)}{\npl(\Tnew)}=\left(t_{\rm BE}x-\mu\right)\frac{e^x}{e^x-1}.
\end{align}

The $\mu$-type spectrum resulting from the
average of two blackbodies is also shown in Figs. \ref{yfig} and
\ref{tefffig}. An important point to note here is that  the Bose-Einstein spectrum is
 uniquely fixed by energy density and number density constraints. In
 particular, the value of the chemical potential $\mu$ is independent of any
 reference temperature we may choose to define the dimensionless variable
 $x$. This is, of course, the same value of $\mu$ we would get if we just comptonize the
 $y$-type distortion of the previous section. The Bose-Einstein spectrum
 and the chemical potential has however more fundamental origins in
 statistical physics, compared to the $y$-type distortion, the shape of
 which originates in  the Compton scattering process or sum of Planck spectra. The $\mu$-type
 results of this section can thus be considered as the basis for our
 definitions of the spectral distortion as pure redistribution of photons,
 and in particular of the $2:1$ division of initial energy in temperature
 perturbations into a blackbody part and a spectral distortion
 part. Further justification is provided by the fact that with this
 definition, the spectral distortion can also be identified with the
 entropy production.

One important difference  from the heating of CMB usually considered,
for example, in clusters of galaxies or due to decay of particles in the
early Universe, is that in mixing of blackbodies, we are also adding
photons (compared to a blackbody at initial average temperature). The additional photons are able to create a new blackbody at a
higher temperature, which is impossible if the photon number density is kept
constant.

Although we have derived the above formulae by considering only
two blackbodies, they are applicable to an ensemble with arbitrary number
of blackbodies by just replacing the average over two blackbodies $(\Delta T/T)^2$ with the average over the
ensemble , $\left < (\Delta T/T)^2\right >$.

\section{Example: Mixing of the blackbodies in the observed CMB sky}
One obvious and cleanest source of blackbodies of different temperatures is
the CMB sky. The temperature in different directions in the sky differs by
$\sim 10\mu {\rm K}$ \citep{cobe2,wmap7} corresponding to the $10^{-5}$ spatial fluctuations in the energy
density of radiation/matter in the early Universe before
recombination\footnote{On very small scales the fluctuations are considerably
smaller since these fluctuations were erased due to diffusion of photons
before and during recombination (Silk damping).}. Thus any telescope, due to finite width
of its beam, looking at the
microwave sky  will inevitably mix the spectra of blackbodies of different
temperature within the beam \citep{cs2004}. In addition we may explicitly average the
intensity over the whole sky to achieve higher sensitivity and precision in
the measurement of CMB spectrum as is done, for
example, by COBE \citep{cobe} and in the proposed experiment PIXIE
\citep{pixie}. The (angular) average amount of energy in CMB anisotropies  is
given by \citep{cs2004},
\begin{align}
\frac{\Delta E}{\Eg}&=6\left<\left(\frac{\Delta
      T}{T}\right)^2\right>=\frac{6}{4\pi}\sum_{\ell=2}^{\infty}(2\ell+1)C_{\ell}\approx 9.6\times 10^{-9},
\end{align}
where $\Eg$ is the energy density of CMB photons.
One third of this energy creates a $y$-distortion of magnitude $y_Y=\frac{1}{2}\left<\left(\frac{\Delta
      T}{T}\right)^2\right>=\frac{1}{12}\frac{\Delta E}{\Eg}=8\times
10^{-10}$. The measured temperature from averaged intensity is also higher
than the averaged temperature $\TCMB$ by $\TCMB\left<\left(\frac{\Delta
      T}{T}\right)^2\right>=4.4{\rm nK}$ accounting for the  remaining  $2/3$ of
energy in anisotropies. The increase in entropy in this mixing is
$\SCMB\times \frac{3}{2}\left<\left(\frac{\Delta
      T}{T}\right)^2\right>=3.5\times 10^{-6}/{\rm cm^3}$.

In the above estimate we ignored the dipole which has been measured by COBE
and WMAP
 \citep{cobe2,wmapdipole} to have an amplitude equal to $3.355 \pm 0.008$ mK corresponding to our peculiar motion
 $v=1.23\times 10^{-3}$. The average power from dipole is then
 $3C_1/(4\pi)=v^2/3$. The resulting $y$ distortion is $v^2/6=2.5\times
 10^{-7}$ and increase in monopole temperature is $v^2/3=1.4\mu {\rm
   K}$. The increase in entropy from the mixing of the CMB dipole  on our sky is $\Delta
 S=1.1\times 10^{-3}/{\rm cm^3}$.

\section{Application:Dissipation of sound waves in the early Universe}
Before recombination, we have a tightly coupled plasma of
radiation-electrons-baryons in the early Universe. At high redshifts, both
the energy density and pressure in the plasma are dominated by radiation
while at low redshifts, but before recombination, the baryon energy density
becomes important, although pressure is still dominated by radiation. Sound
speed in this relativistic plasma is therefore $1/\sqrt{3}$ and Jeans scale
or sound horizon is   particle horizon$/\sqrt{3}$. Primordial
perturbation on scales smaller
than the Jeans scale or sound horizon therefore oscillate setting up
standing sound waves \citep[][see also \citealp{sz1970b}]{lifshitz}. Although the photon mean free path due to Compton
scattering on electrons is very small, they are still able to traverse considerable
distance since the big bang, performing a random walk among the electrons. This
diffusion and mixing of photons as a result of Thomson scattering erases the sound waves on scales corresponding to
the diffusion scale (and smaller). Macroscopically, the dissipation of
sound waves
can be identified as due to the \emph{shear viscosity} and \emph{thermal conduction} in
the relativistic fluid composed of baryons, electrons and
photons. The damping of sound waves on small scales due to thermal
conduction was pointed out by \citet{lifshitz} and first calculated by
\citep{silk} and is known as Silk damping. At high redshifts ($z\gg 693$) when the
energy density of the plasma is also dominated by radiation, shear
viscosity is more important than thermal conductivity and was calculated by
\citet{Peebles1970}, later \citet{kaiser}  included the effect of
polarization \citep[see also][]{Weinberg1971}. The resulting spectral
distortions in CMB were considered by
\citet{sz1970c,Daly1991,hss94}. \citet{sz1970c} demonstrated that the upper
limit to the $\mu$-type distortions allows us to constrain the amplitude of
the primordial fluctuations, which were completely damped in the CMB (on small scales), and are today
in the
unobservable part of the matter/CMB power spectrum.

\begin{figure}
\resizebox{\hsize}{!}{\includegraphics{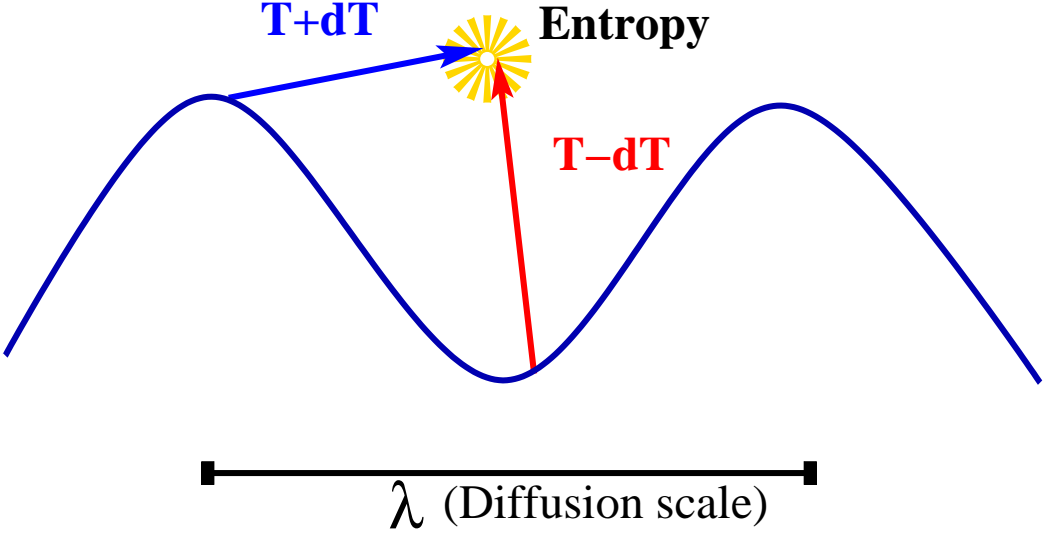}}\caption{\label{schematic}
  Cartoon picture of mixing of
    blackbodies: Photons from different phases of sound wave mix and create spectral distortion/entropy.}
\end{figure}

Microscopically, diffusion of photons mixes photons from different phases of
sound waves which have different temperatures. This is shown schematically
in Fig. \ref{schematic}. This implies that locally a $y$-type
distortion is created, which quickly comptonizes to a $\mu$-type distortion
at $z\gtrsim 10^5$, as calculated in the previous sections. The dissipation of sound
waves is best understood in Fourier space, denoting Fourier transform of
$\frac{\Delta T}{T}(\mathbf{x},\mathbf{\hat{n}})$ with
$\Theta(\mathbf{k},\mathbf{\hat{n}})$, $\mathbf{\hat{n}}$ is the photon
direction\footnote{Bold letters denote vectors, bold letters with hat
  denote unit vectors and normal letters denote magnitude of the vector}, $\mathbf{x}$ is the comoving coordinate and
$\mathbf{k}$ is the comoving wavenumber \citep[see also][]{cks2012},
\begin{align}
\left.\frac{\Delta E}{\Eg}\right|_{\rm acoustic}&=6\left<\left(\frac{\Delta
      T}{T}\right)^2\right>\nonumber\\
&=6\int \frac{\id^3 k}{(2\pi)^3}e^{i\mathbf{k.x}}\int \frac{\id^3 k'}{(2\pi)^3}\left<\Theta(\mathbf{k'},\mathbf{\hat{n}})\Theta(\mathbf{k-k'},\mathbf{\hat{n}})\right>\nonumber\\
&=6\int \frac{k^2\id k}{2\pi^2}P_i(k)\left[\sum_{\ell=0}^{\infty}(2\ell+1)\Theta_{\ell}^2(k)\right]\label{tderiveq},
\end{align}
where we have expanded the temperature perturbation transfer functions in Legendre polynomial
basis,
$\Theta(\mathbf{\hat{n}.{\hat{k}}},k)=\sum_{\ell}(-i)^{\ell}(2\ell+1)\mathcal{P_{\ell}}(\mathbf{\hat{n}.{\hat{k}}})\Theta_{\ell}(k)$,
$\mathbf{\hat{n}}$ is the photon direction, $\mathbf{k}$ is the unit vector
along the Fourier mode which is parallel to the electron peculiar velocity
in linear theory. This transformation is possible since in first order
perturbation theory the photon transfer functions depend only on
$\mathbf{\hat{n}.{\hat{v}}}$ or in Fourier space on
$\mathbf{\hat{n}.{\hat{k}}}$ and not on $\mathbf{\hat{n}}$ and
$\mathbf{\hat{k}}$ separately. $P_i(k)$ is the power spectrum of initial
curvature perturbations ($\zeta_i$) with respect to which the transfer functions
$\Theta_{\ell}$ are calculated. We have also used homogeneity and isotropy
of the Universe
to carry out one of the integrals and angular part of the second integral.
Cosmological perturbations, in particular monopole and dipole depend on the
choice of gauge. We will use conformal Newtonian gauge from now on. The
spectral distortions defined as pure redistribution of photons, for example
the $y$ distortions we are considering, are however gauge independent. 

Before cosmological recombination starts with helium recombination at
$z\sim 6000$, electrons/baryons and radiation are tightly coupled and
$\ell>2$ modes can be neglected. Most of the energy of sound waves is in
monopole and dipole and can be calculated using relation between monopole
and dipole in the tight coupling regime, $\Theta_1\approx
\Theta_0/\sqrt{3}$ (using sound speed $c_s\approx 1/\sqrt{3})$, 
\begin{align}
\left.\frac{\Delta E}{\Eg}\right|_{\rm acoustic}&=6\int \frac{k^2\id k}{2\pi^2}P_i(k)\left[\Theta_{0}^2+3\Theta_{1}^2\right]\nonumber\\
&\approx \int \frac{k^2\id
  k}{2\pi^2}P_i(k)\left[12\Theta_{0}^2\right]
\end{align}
This result is $9/4$ times the estimate used in
\citet{sz1970c,hss94,ksc2011} where it was also assumed that all of the
energy in sound waves gives rise to spectral distortions. As discussed above, $1/3$ of this energy,
when dissipated due to mixing of blackbodies,
sources spectral distortions which are created as $y$-type but rapidly
comptonize to $\mu$-type distortions or Bose-Einstein spectrum at high
redshifts $z\gtrsim 10^5$. The remaining $2/3$ of the dissipated energy raises the average
temperature of CMB which is not directly observable. Thus the correct
result for distortions differs from earlier estimates only by a factor of
$1/3\times 9/4=3/4$ \citep{cks2012}.

At redshifts $10^5 \lesssim z\lesssim \pot{2}{6}$, the average $\mu$ distortion
therefore increases at a rate,
\begin{align}
\frac{\id\mu}{\id t}&=-1.4\frac{\id}{\id t}\frac{1}{3}\left.\frac{\Delta
  E}{\Eg}\right|_{\rm acoustic}\nonumber\\
&\approx-\frac{\id}{\id t} \int \frac{k^2\id
  k}{2\pi^2}P_i(k)\left[5.6\Theta_{0}^2\right]\label{monoeq}
\end{align}
and rate of increase of entropy density is given by,
\begin{align}
\frac{\id\Delta S/S}{\id t}&=-\frac{1}{4}\frac{\id}{\id t}\left.\frac{\Delta
  E}{\Eg}\right|_{\rm acoustic}\nonumber\\
&\approx-\frac{\id}{\id t} \int \frac{k^2\id
  k}{2\pi^2}P_i(k)\left[3\Theta_{0}^2\right].
\end{align}
The above rates are easily calculated using analytic tight coupling
solutions given by \citet{hu1995} and for power spectrum with constant
scalar index it is possible to do the time derivatives and integral
analytically. In particular the expressions presented in \citet{ksc2011}
for $\mu$ type distortions remain valid after multiplication by a factor of
$3/4$. The $y$- type distortions at $z\lesssim 10^4$ require inclusion of additional modifications due to
breakdown of tight coupling during recombination, second order Doppler effect, and higher order temperature anisotropies, which were derived in \citet{cks2012} using second order Boltzmann equation.

 We can  use
the first order Boltzmann equation,
$\id\Theta / \id t = \Theta_0-\frac{1}{2}[\Theta_2+\Theta^{\rm
  P}_0+\Theta^{\rm P}_2] \mathcal{P}_2(\mathbf{\hat{n}.{\hat{k}}})-
\Theta(\mathbf{\hat{n}.{\hat{k}}}) - i {\rm v} \,\mathcal{P}_1(\mathbf{\hat{n}.{\hat{k}}})$, to calculate the time derivative of
Eq.~\eqref{tderiveq}. Taking into account that only 1/3 of dissipated energy leads to
spectral distortions, and requiring that the dipole/velocity term is gauge
invariant gives us the full result,\footnote{We have ignored the
  gravitational potential/metric perturbations since they
  cannot create spectral distortions. 
They do cancel out explicitly in the second order Boltzmann equation
\citep{cks2012}. Gravitational potential/metric perturbations do contribute
to 
the average CMB temperature but this effect is unobservable.}
\begin{align}
\frac{\id}{\id t}\left.\frac{\Delta E}{\Eg}\right|_{\rm
  distortion}=&-4\left<\frac{\Delta T}{T}
\frac{\id}{\id t}\frac{\Delta T}{T}\right>&\nonumber\\
\xrightarrow{\rm ignore~ metric~ perturbations~}~&4 \Ne\sigT
 \int \frac{k^2\id
  k}{2\pi^2}P_i(k)\left[\Theta_1\left(3\Theta_1-\ve\right)\right.&\nonumber\\
+\frac{9}{2}\Theta_2^2-&\left.\frac{1}{2}\Theta_2\left(\Theta_2^{\rm P}+\Theta_0^{\rm P}\right)+\sum_{\ell\ge 3}(2\ell+1)\Theta_{\ell}^2\right]&\nonumber\\
\xrightarrow{\rm impose~ gauge~ invariance~}~&4 \Ne\sigT
 \int \frac{k^2\id
  k}{2\pi^2}P_i(k)\left[\frac{\left(3\Theta_1-\ve\right)^2}{3}\right.&\nonumber\\
+\frac{9}{2}\Theta_2^2-&\left.\frac{1}{2}\Theta_2\left(\Theta_2^{\rm P}+\Theta_0^{\rm P}\right)+\sum_{\ell\ge 3}(2\ell+1)\Theta_{\ell}^2\right],&\label{finaleq}
\end{align}
where $\mathbf{v_{\rm e}}(k)\equiv -i\hat{\mathbf{k}}\ve(k)$ is the
transfer function of peculiar velocity of baryons/electrons and
$\Theta_\ell^{\rm P}$ denote polarization multipole moments. 
This expression  was first derived by \citet{cks2012} using the second order Boltzmann
equation, which automatically takes care of the gauge independence and metric
perturbations. Using that the spectral distortions (defined as
a pure redistribution of photons) are gauge invariant, together with the
fact that the only physical mechanism operating here is the mixing of
blackbodies, allows us to derive the full result without referring to
the second order Boltzmann equation and using only the well studied first
order Boltzmann equation. The identification of a symmetry in the problem,
i.e. gauge invariance, allows us to derive very simply    the
 results of the extensive calculation of \citet{cks2012} corresponding to the  average spectral
 distortions in CMB created by the dissipation of sound waves. Using the
 Boltzmann equation, on the other hand, also allows \citet{cks2012} to
 obtain new results on anisotropies  of the spectral distortions, and we
 refer to that work for a detailed discussion.

We note that with our definitions, all the photon
transfer functions and $v$ are real quantities. We have also the introduced multipole moments of
degree of polarization, $\Theta^{\rm P}$, defined in the same way as
the temperature multipole moments.\footnote{Our convention is same as that
  of \citep{mabert95,dod} but differs from
  that of \citet{zh95} by a factor of $(-i)^{\ell}$ in the definition of
  multipole moments.} The
polarization terms are coming directly from the first order  Boltzmann
equation for temperature. They contribute at the level of $\sim 5\%-10\%$ to the effective heating rate close to the recombination epoch at $z\simeq 10^3$ \citep{cks2012}.

 At lower redshifts, baryons and photons develop relative
velocity and second order Doppler effect also contributes to the $y$
distortions and appears above in the gauge invariant combination with
photon dipole. This effect can thus be considered as the mixing of dipole in the
electron rest frame but in a general frame like conformal Newtonian gauge
it originates in the Compton collision term \citep{hss1994}. This is the
only significant contribution from the second order Compton collision term
to the spectral
distortions (in addition of course to the Kompaneets term). This term can
also be 
easily obtained by taking the part of the second order Compton collision
term \citep[see for example][]{hss1994, Bartolo2007, Pitrou2009} with the $y$-type spectral dependence. Higher order
corrections originating in the terms which are second order in perturbation
theory and also second order in energy transfer were  calculated in
 \citet{cks2012} and shown to be negligible. Fitting formulae for $\mu$
distortions as a function of spectral index and its running for primordial
adiabatic perturbations are also given in \citet{cks2012}. 
Recently  isocurvature modes were considered by \citet{dent2012},
  \citet{ceb2012} have calculated  the distortions from some  exotic models for small-scale power spectrum,
  \citet{pajer2012} have pointed out the possibility of constraining
  non-gaussianity using $\mu$ distortions, and \citet{ganc2012} have
  investigated the consequences of modified initial state for single field inflation.

Equation
\eqref{finaleq} is explicitly gauge invariant and is recommended for
 calculations of distortions instead of taking the time derivative of monopole,
Eq. \eqref{monoeq}. In particular, Eq. \eqref{monoeq} is accurate only in
the $\mu$-era ($z\gtrsim 10^5$), and the early stages of the $y$-era ($z\gtrsim 10^4$), when only the  shear viscosity (quadrupole) term contributes,
and cannot be used to  estimate the $y$-type distortions created at late times, around and after recombination. In the $y$-era 
($z\lesssim 10^4$), thermal conductivity (dipole/velocity) term and $\ell>2$
anisotropies  contribute at a  significant level
 and the full Eq. \eqref{finaleq} must be used.

The first three terms in Eq. \eqref{finaleq} give the dominant contribution to
the dissipation of sound waves. The first term  mixes the blackbodies in
the dipole resulting in transfer of heat along the temperature gradient,
and can thus be identified as the effect of thermal conductivity. The
second term in Eq. \eqref{finaleq}, similarly, mixes the blackbodies in the quadrupole or the shear
stress in the photon fluid and can thus be identified as the
effect of shear viscosity. The third term takes into account the
polarization dependence of the Compton scattering and is a correction to
the shear viscosity  \citep{kaiser}. The $\ell\ge 3$ multipoles are negligible during tight coupling by
definition (and thus for the $\mu$-type distortions) but give a small
contribution during recombination as the tight coupling breaks down and the
photons begin to free stream \citep{ksc2011,cks2012}. On substituting the
conformal Newtonian gauge tight
coupling solutions  \citep{hu1995,zh95,dod}, we get for the first term (ignoring the phase of the
oscillations which actually differs between the left hand side and the
right hand side by $\pi/2$)\footnote{This does not introduce any error in
  the calculation of heating of the average CMB spectrum  since we should
  average the sound wave over a whole oscillation.}, 
\begin{align}
\frac{\left(3\Theta_1-\ve\right)^2}{3}\approx
\frac{R^2}{1+R}\frac{k^2}{(\Ne \sigT)^2}\Theta_1^2,\label{fn2}
\end{align}
where $R=\frac{3 \rho_b}{4E_{\gamma}}=\frac{693}{1+z}$ and $\rho_b$ is the energy density
of baryons. At $z> 10^5$, when the $\mu$-type distortions are created, $R^2\ll
1$ and thermal conductivity  contributes negligibly  to the sound wave dissipation.
 The dominant terms during the  $\mu-$type era are 
  the  second and the third (shear viscosity) terms \citep{zh95} 
\begin{align}
\frac{9}{2}\Theta_2^2-\frac{1}{2}\Theta_2\left(\Theta_2^{\rm
    P}+\Theta_0^{\rm P}\right)&=\frac{16}{15}\frac{k^2}{(\Ne \sigT)^2}\Theta_1^2.\label{fn3}
\end{align}
If we ignore polarization, the factor of $16/15$ in the above equation
would be replaced by $8/9$. 
At redshifts $z\gtrsim 2\times 10^6$ the
distortions are exponentially suppressed due to the combined action of
bremsstrahlung and double Compton, which can create and destroy photons at
low frequencies, and comptonization, which redistributed the photons over
the entire spectrum creating a Bose-Einstein spectrum. The rate of energy
injection in Eq. \eqref{finaleq} therefore has to be multiplied by a suppression factor or blackbody
visibility, $\mathcal{G}(z)$, for $z\gtrsim 10^5$, giving the part of the energy injection which
is actually observed as $\mu$-type distortion. An analytic solution for $\mathcal{G}(z)$ was
calculated by \citet{sz1970}, who only considered bremsstrahlung. Their
solution was  later
applied to double Compton scattering by \citet{dd1982} (accurate to $5-10\%$) and improved recently
to sub-percent accuracy in \citet{ks2012}.
Numerical computation of  the spectral distortions is possible using numerical codes such as {\sc
  KYPRIX} \citep{pb2009} and {\sc
  CosmoTherm}\footnote{www.chluba.de/CosmoTherm} \citep{cs2011}, the later
code includes the energy injection due to Silk damping and is able to
calculate such small distortions at high precision.

\section{Conclusions}
Mixing of blackbody spectra results in a photon distribution which is no longer a perfect  blackbody but
contains a $y$-type spectral distortion. The mixed spectrum  has higher
entropy than the average entropy of initial spectra as expected from an
irreversible process which creates disorder. The energy which goes into the
spectral distortion and the increase of entropy can be calculated very
simply using statistical physics. The increase
in entropy is, at first order in small distortions, independent of the
shape of the distortion and can be calculated using the equilibrium
thermodynamics formula for small addition of heat, $dS=dQ/T$. The part of
the energy
which sources distortions is only  $1/3$ of the total  energy
available in temperature perturbations
(with respect to a blackbody at average initial temperature) and also
results in increase of entropy. The remaining $2/3$ of the
energy in perturbations goes into increasing the average blackbody
temperature of the photons and can be identified with entropy conserving
part of the mixing process. We have proven explicitly in this
  paper that the comptonization of the initial spectrum with $y$-type
  distortion to the Bose-Einstein spectrum does not
  change this 2:1 division of the dissipated energy into a blackbody part
  and a distortion part. From an observational point
  of view, we are just interested in the value of the observable $\mu$, and
  this 2:1 division of dissipated energy allows us to compute the value of $\mu$ in a 
  straightforward way. $\mu$-distortions are unique and very important
  because it is impossible to create them after $z\lesssim 10^5$ and thus
  probe the physics of the early Universe unambiguously. $y$-type distortions on the
  other hand are created throughout the later history of the Universe, and
it is difficult to separate the contributions from the different epochs.

We have an almost perfect blackbody in the Universe in the form of CMB. We
apply our results to mixing of blackbodies in the observed CMB sky. As a
result of this mixing, the spectrum observed by a telescope with finite
beam size would have inevitable $y$-type distortions. This effect must be
taken into account in experiments aiming to measure CMB spectrum at high
precision. A very important application of physics of the mixing
blackbodies is in the early Universe. Before recombination, the tightly
coupled baryon-electron-photon plasma is excited by primordial
perturbations in energy density, resulting in standing sound waves. The
photons in different phases of the sound wave have a blackbody spectrum
with different temperature. Photon diffusion and isotropization of
  the radiation field
  by
  Thomson scattering mixes these blackbodies on
scales corresponding to diffusion length. These spectral distortions measure the
primordial spectrum on very small scales (with the smallest scales
completely inaccessible by any other means), at comoving wavenumbers $10
\lesssim k \lesssim 10^4~ {\rm Mpc}^{-1}$, and will thus provide a powerful
new tool to constrain early Universe physics in the future.
We have derived the  energy
release resulting from the  damping of these sound waves, and the corresponding spectral
distortions of the CMB, in a  simple manner using the physics of mixing of
blackbodies. These results and additional (but negligible) corrections were
calculated recently using second order perturbation theory in \citet{cks2012}.
The results, for the very important $\mu$ type distortions, coincidentally
are close to the
estimates used in literature until now, with the main difference being a
correction  factor of $3/4$. 

\bibliographystyle{aa}
\bibliography{bbmixing}
\end{document}